\documentclass[conference]{IEEEtran}
\IEEEoverridecommandlockouts

\usepackage{cite}
\usepackage{amsmath,amssymb,amsfonts}
\usepackage{algorithmic}
\usepackage{graphicx}
\usepackage{textcomp}
\usepackage{xcolor}
\def\BibTeX{{\rm B\kern-.05em{\sc i\kern-.025em b}\kern-.08em
    T\kern-.1667em\lower.7ex\hbox{E}\kern-.125emX}}
\begin{document}

\title{Quantized AI Inference on Constrained Embedded Platforms for Small-Satellite Settings\\
}

\author{\IEEEauthorblockN{Carlos Rafael Tordoya Taquichiri}
\IEEEauthorblockA{\textit{Institute of Embedded Systems} \\
\textit{ZHAW School of Engineering}\\
Winterthur, Switzerland \\
tord@zhaw.ch}
\and
\IEEEauthorblockN{Hans Dermot Doran}
\IEEEauthorblockA{\textit{Institute of Embedded Systems} \\
\textit{ZHAW School of Engineering}\\
Winterthur, Switzerland \\
donn@zhaw.ch}
\and
\IEEEauthorblockN{Pablo Ghiglino}
\IEEEauthorblockA{\textit{R\&D Department} \\
\textit{Klepsydra Technologies}\\
Zurich, Switzerland \\
pablo.ghiglino@klepsydra.com}
}

\maketitle

\begin{abstract}
In resource-constrained small-satellite settings, AI inference must operate under tight size, power, and payload budgets, which tend to limit onboard compute capability and data handling. These conditions motivate establishing a clear baseline for quantized AI inference under bounded compute and memory resources. To instantiate this baseline, a representative embedded-vision neural-network workload serves as the reference case.
With this motivation, this paper presents a measurement-based characterization of quantized execution for this AI workload on highly constrained embedded platforms (for instance, Cortex-M), grounded as a lower-bound operating point. In this regime, scaling tends to rely on explicit orchestration rather than OS-managed, transparent multicore scheduling, and timing behavior is shaped by instruction efficiency and memory movement. As a result, the characterization provides a structured reference for estimating execution time across orchestrated configurations (e.g., multiple cores and/or devices), treating orchestration and architectural variation as explicit design choices.
We report latency metrics alongside data-movement observations, and interpret these measurements in light of ALU/SIMD utilization under quantized arithmetic for the Cortex-M. Finally, we outline how this baseline provides a reference point for positioning the results against more space-typical embedded processor classes (e.g., LEON/NOEL-V).
\end{abstract}


\section{Introduction}
\subsection{Motivation}

Onboard artificial intelligence is becoming increasingly relevant for small-satellite and space-oriented embedded systems, where local data processing reduces dependence on ground communication links, decreases decision latency, and supports more autonomous operation. This is particularly important when sensing payloads generate more data than what can be continuously downlinked, or when timely decisions must be made under limited connectivity and strict mission constraints \cite{b1}.

However, executing AI inference in this context is constrained by tight limits on power consumption, memory capacity, processor performance, payload size, and thermal budget. These restrictions are especially significant for small-satellite platforms, where the available computing resources tend to be below those of terrestrial edge-AI devices. In addition, space-oriented processing platforms often prioritize reliability and determinism over raw computational throughput. As a result, deploying modern neural-network workloads directly on such systems remains challenging, particularly for operations such as convolutions and matrix multiplications, which constitute the major computational components in embedded-vision models \cite{b2}.

To make neural-network inference feasible under these constraints, models are commonly quantized to reduce arithmetic and memory requirements by replacing floating-point operations with low-precision integer arithmetic. However, the performance of quantized inference on highly constrained embedded processors is not determined only by the number of arithmetic operations. It is also strongly affected by instruction-level efficiency, DSP/SIMD utilization, data layout, memory movement, and AI framework overhead. In addition, more complex multicore, multi-processor, or accelerator-based deployments can improve the overall AI inference performance, but they also introduce communication, synchronization, scheduling, data-movement and power-consumption considerations. Implementing such configurations therefore requires additional effort and does not automatically guarantee a meaningful speedup. Their benefit depends on whether the expected acceleration is larger than the overheads introduced by orchestration and data movement. For this reason, a measured execution baseline is needed to quantify the local execution cost of a representative quantized workload independently of inter-processor data movement, before evaluating additional optimization or offloading strategies.

Motivated by these constraints, this work characterizes quantized convolution execution on a highly resource constrained platform as a lower-bound operating point for embedded AI inference. The goal is to establish an execution reference for a representative quantized convolution workload. This reference is then used to interpret scalar execution, DSP/SIMD-optimized microkernels, scalability and communication overhead in the context of orchestrated configurations involving multiple cores, space-oriented processors such as LEON or NOEL-V, or embedded FPGA-based acceleration.

\subsection{Proposed Solution}

To address the need for a reference execution baseline, this work presents a measurement-based characterization of quantized neural-network execution on a highly constrained embedded platform. The paper establishes the local execution cost of a representative quantized convolution layer extracted from an embedded-vision workload. This provides a lower-bound reference for the core computation under bounded compute and memory resources.

The characterization focuses on the execution behavior of the quantized math backend employed by the Klepsydra AI framework. The selected convolution workload is executed using two paths of this backend: a scalar/ALU implementation and a DSP/SIMD-optimized implementation. The scalar path provides the constrained local reference, while the DSP/SIMD path shows the effect of exploiting the instruction-level capabilities of the Cortex-M core. Since both paths execute the same workload, the comparison makes the impact of hardware-oriented backend optimization easier to interpret.

In this work, the characterization is performed in a Zephyr-based environment, chosen for its POSIX support. The same selected workload is also executed using TensorFlow Lite Micro \cite{tflite_micro} as an external embedded-inference reference. This comparison shows how backend implementation affects the execution of the same low-level quantized layers under constrained embedded conditions.

In addition to local convolution latency, OpenAMP communication overhead between two Cortex-M cores is measured independently. These measurements provide a first-order estimate of the communication and synchronization cost that orchestrated configurations must consider. Combined with the local execution baseline, it supports an early feasibility check for offloading computation to another core, processor, or accelerator.

Finally, the baseline can be used to calculate the expected timing behavior for complex orchestration implementations. This provides an early estimation of the potential execution improvement in such configuration and also provides a plausibility reference for checking whether the implementation behaves as expected once it is implemented.

\section{Related Work}

\subsection{On-board AI and Space-Oriented Embedded Processing}

Prior in-orbit demonstrations have shown the practical value of moving AI inference closer to the sensing payload. ESA's $\Phi$-Sat-1 demonstrated onboard neural-network inference for Earth-observation data filtering, reducing the need to downlink unusable imagery \cite{b3}. Such mission-level results support the relevance of onboard AI, while the resource constraints reported for such platforms motivate a closer measurement-based understanding of execution time, memory footprint, and energy-efficiency during AI inference.

The feasibility of such processing strongly depends on the available on-board computing platform. Space-oriented processors are selected not only for performance, but also for critical properties such as reliability, power consumption, and radiation tolerance. Among space-oriented processing technologies, LEON-based devices such as the GR740~\cite{b4} represent radiation-hardened multicore SoC designs, while NOEL-V represents the move toward configurable RISC-V processor cores for space-grade systems~\cite{b5}. In addition, radiation-tolerant reconfigurable devices, including RTG4~\cite{b6}, RT PolarFire~\cite{b7}, and space-grade Versal adaptive SoCs~\cite{b8}, form part of the hardware design space for data-intensive payload processing and acceleration.

These platforms make heterogeneous execution attractive, but also introduce communication, synchronization, and data-movement costs. A constrained local execution baseline is therefore useful for estimating the potential timing benefit of orchestration before implementation, and for further evaluation whether multicore or heterogeneous execution introduces unexpected overheads, imbalance, or bottlenecks.

\subsection{Quantized Inference and Embedded Backends}

Quantized inference is a common approach for executing neural networks on highly constrained embedded devices. By representing weights and activations with low-precision integer values, quantization reduces memory footprint and enables inference without floating-point hardware. Integer-only quantization has been shown to preserve useful model accuracy while enabling efficient execution on embedded processors~\cite{b9}. These properties make quantized inference a natural operating point for edge and on-board space applications, where memory capacity and processor resources are constrained.

However, quantization alone does not guarantee low latency. The execution time of a quantized convolution layer depends strongly on the arithmetic backend, data layout, and the ability to exploit architecture-specific instructions. In this domain, TensorFlow Lite Micro provides a widely used embedded-inference runtime for microcontroller-class systems~\cite{b10}. It can be used with CMSIS-NN, which provides Cortex-M-optimized kernels that exploit DSP capabilities~\cite{b11}.

\subsection{Runtime Support and Explicit Orchestration}

Embedded AI performance is affected not only by the neural-network model and quantization scheme, but also by the runtime framework used to invoke operations and select backend implementations. This is relevant for constrained and heterogeneous systems, where potential scaling may rely on explicit orchestration across cores, processors, or accelerators rather than transparent operating-system scheduling.

Klepsydra AI is relevant in this context because it provides a highly scalable lock-free execution model in which backend-specific implementations can be integrated and compared within a common framework structure~\cite{b12}. Such a framework supports explicit orchestration when dealing with AMP platforms or offloading AI operations to accelerators. Prior work has explored Klepsydra AI for onboard inference with dedicated co-processing backends in space-oriented applications~\cite{b13}.

Explicit orchestration also introduces communication and synchronization costs. OpenAMP addresses asymmetric multiprocessing systems through inter-processor communication mechanisms such as RPMsg~\cite{b14}. For this reason, the local backend measurements in this paper are complemented with OpenAMP communication measurements, providing a first reference for reasoning about future offloaded or heterogeneous configurations.

\section{Baseline Characterization Approach}

This work defines a baseline-oriented characterization methodology to quantify the cost of AI execution before executing complete end-to-end models or before introducing explicit workload orchestration or accelerator offloading. The objective is to establish a measured reference point for a representative quantized convolution workload under constrained embedded conditions, which can later support latency estimation for complete inference models and more complex orchestrated configurations.

\subsection{Baseline Scope}

Complete neural-network inference includes effects such as layer scheduling, inter-layer buffering, preprocessing, postprocessing, and application-level control flow. These effects are relevant for complete deployments, but they can obscure the latency measurement of quantized arithmetic in AI operations. By isolating one convolution layer, the characterization focuses on the execution cost of the same workload under different implementation paths.

Therefore, in this work, the baseline workload is defined as a representative quantized convolution layer from a MobileNetV2 embedded-vision model. The scalar/ALU execution of this layer on the Cortex-M33 platform is treated as the constrained baseline.

To define a clear baseline with limited system-level variability, this work uses a highly constrained low-power embedded platform as the base evaluation target. This is done to avoid the influence of complex operating-system services, dynamic scheduling behavior, and heterogeneous runtime effects that may obscure the cost of the selected quantized AI workload. Accordingly, the selected platform is an Arm Cortex-M33 platform running Zephyr, which is a representative case in which compute resources, memory capacity, instruction support, and communication mechanisms are explicitly constrained.

\subsection{Measured Execution Paths and Quantities}

Since the characterization focuses on execution behavior, the primary metric is the latency of the selected quantized convolution layer. As mentioned before, the measured execution paths include the scalar/ALU path of the characterized quantized math backend, used as the constrained baseline, and the DSP/SIMD-optimized path of the same backend, used to quantify the benefit of architecture-aware backend optimization under the same workload and platform conditions. In addition, TensorFlow Lite Micro reference executions are performed for the same workload, using its scalar implementation and its CMSIS-NN optimized path for benchmarking.

\subsection{Baseline in Orchestration Reasoning}

Beyond direct benchmarking, the measured latency can also be converted into a backend-specific timing indicator for early latency estimation. This is possible because convolution has a well-defined arithmetic workload where, for a given layer configuration, the theoretical number of multiply-accumulate (MAC) operations can be computed from parameters such as input and output dimensions, kernel size, channel count, stride, and padding. Thus, the measured latency can be related to the amount of computation executed in a layer by a specific AI math backend. This relation defines the timing indicator, which can be expressed as:

\begin{equation}
\kappa_b = \frac{T_{\mathrm{conv},b}}{N_{\mathrm{MAC}}},
\end{equation}

where \(T_{\mathrm{conv},b}\) is the measured execution time of the selected convolution using backend \(b\), and \(N_{\mathrm{MAC}}\) is the number of MAC operations required by the layer. The resulting quantity \(\kappa_b\) represents an empirical time-per-MAC for a specific platform and backend implementation. In this work, \(\kappa_{\mathrm{ALU}}\) is derived from the scalar/ALU baseline, while the DSP/SIMD-optimized and TensorFlow Lite Micro paths provide additional indicators for backend and framework comparison.

This indicator captures the practical behavior of the selected execution path, including backend implementation efficiency, instruction use, memory-access behavior, and data layout. At the same time, because the measurement is obtained from an isolated convolution workload on a constrained embedded platform, it excludes additional effects from complete model execution, application-level control flow, and explicit offloading mechanisms.

As a first-order estimate, the latency of another convolution layer executed using the same backend/framework can be approximated as

\begin{equation}
\hat{T}_{l,b} = \kappa_b * N_{\mathrm{MAC},l},
\end{equation}

where \(N_{\mathrm{MAC},l}\) is the MAC count of layer \(l\). This estimate is meaningful for layers with similar operation types and quantization format. Therefore, it provides an early indication of whether a model, layer group, or backend choice is likely to satisfy a target latency budget before implementing a complete inference pipeline.

\begin{figure}[htbp]
\centerline{\scalebox{0.20}{\includegraphics{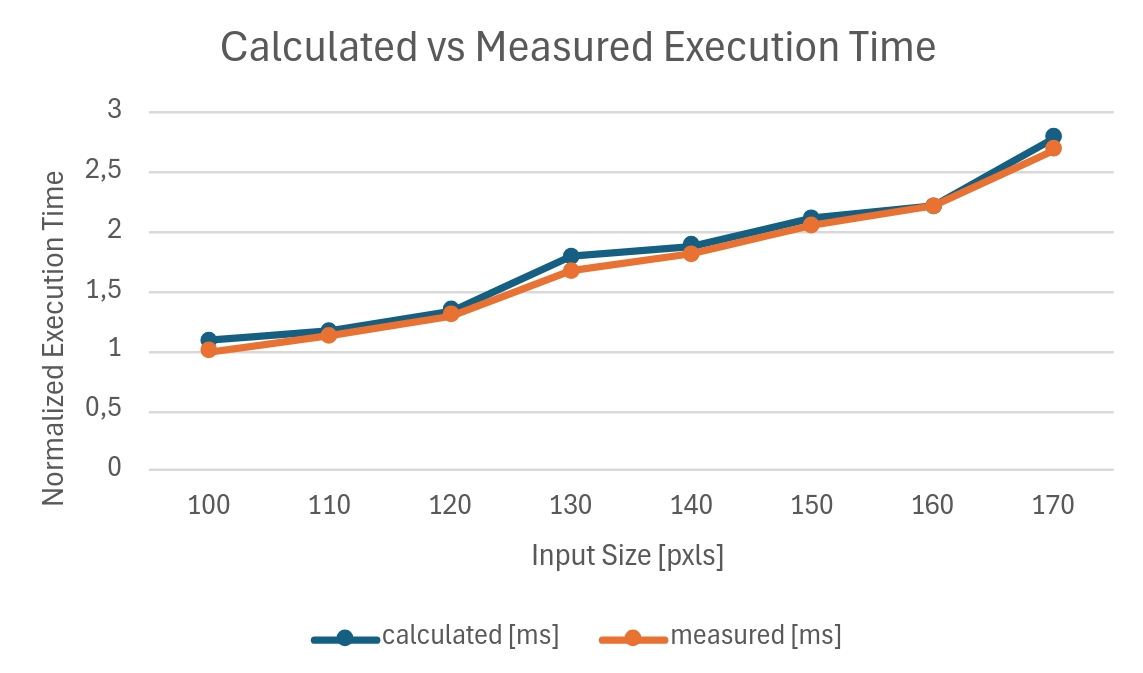}}}
\caption{Comparison of estimated and measured execution times for YOLOv8n on a typical embedded platform (XMOS.ai)}
\label{fig_macs}
\end{figure}

A prior application of this MAC estimation methodology showed close agreement between estimated and measured execution time across several input shapes when the operation type and execution path were kept fixed, as illustrated in Fig.~\ref{fig_macs}. This supports its use here as a planning-level estimator.

For explicitly orchestrated execution, the estimated computation time must be considered together with communication and synchronization overhead. In this work, the OpenAMP measurements provide representative inter-core overheads for the Cortex-M33 platform. For a simple blocking offload, where the local processor waits for the remote result, the offloaded execution time can be approximated as

\begin{equation}
\hat{T}_{\mathrm{offload}} =
\hat{T}_{\mathrm{remote}} + T_{\mathrm{comm}},
\end{equation}

where \(\hat{T}_{\mathrm{remote}}\) is the estimated execution time on the remote processing element, and \(T_{\mathrm{comm}}\) represents the communication and synchronization overhead for the selected interaction pattern. In this case, offloading is beneficial only if the combined remote execution and communication cost is lower than the corresponding local execution time or if the local processor is busy during offloading operation.

In this way, the baseline and the \(\kappa_b\) indicators support the estimation of computation time, enable comparison with communication overhead, and provide a first KPI to assess whether a multicore, multi-processor, or accelerator-based implementation is expected to provide a net timing benefit before implementing the full system.

\subsection{Model-Level Timing Estimation}

\begin{figure}[htbp]
\centerline{\scalebox{0.22}{\includegraphics{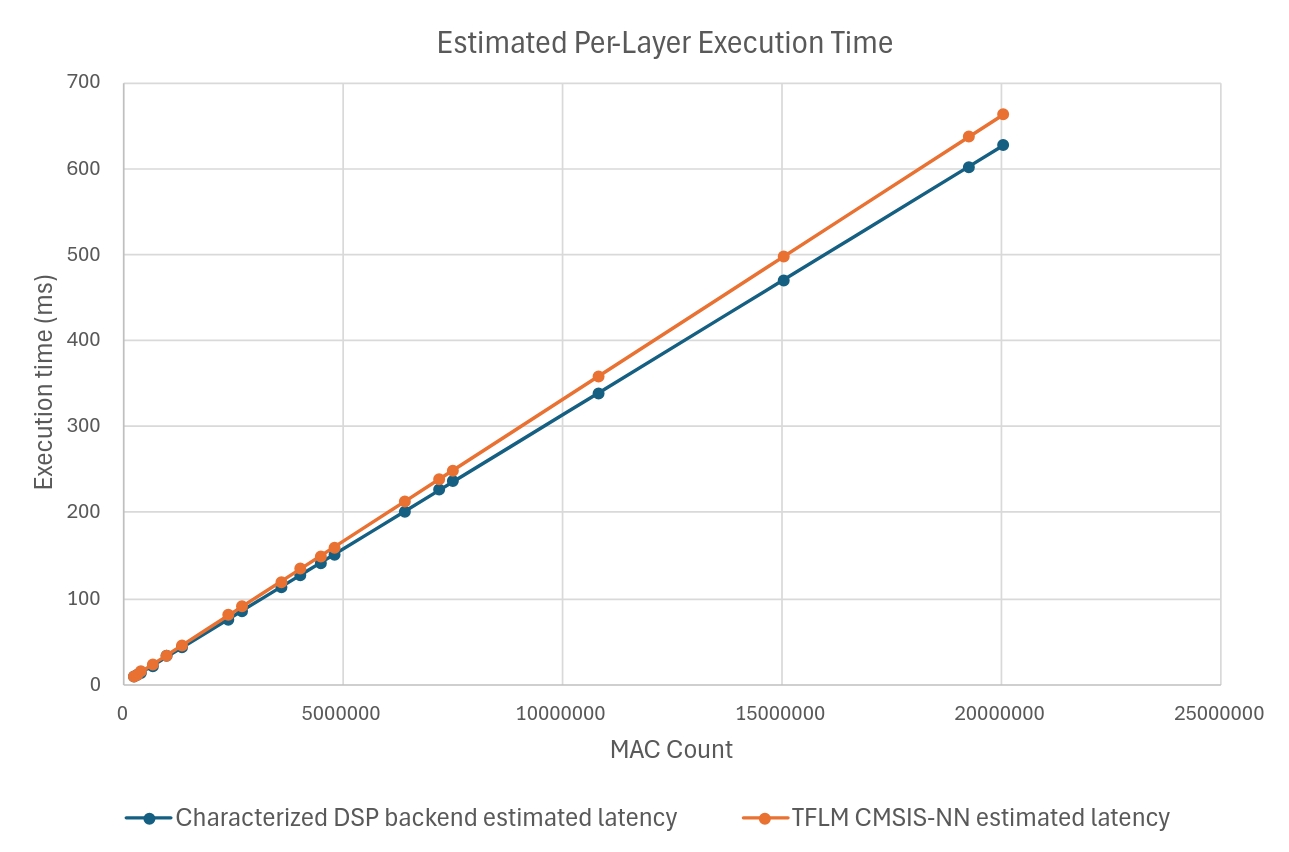}}}
\caption{Estimated execution time per MobileNetV2 layer on the Cortex-M33 platform.}
\label{fig_kpsr_tflm}
\end{figure}

For the Cortex-M33 platform, this estimation methodology provides a first indication of how latency scales from individual convolution layers toward full AI model execution. For MobileNetV2, the resulting values are on the order of \(9.35\,\mathrm{s}\) for the DSP path of the characterized backend and \(9.88\,\mathrm{s}\) for the DSP-optimized TensorFlow Lite Micro path using CMSIS-NN. Fig.~\ref{fig_kpsr_tflm} shows the corresponding layer-level estimates used to obtain these model-level timing indications.

\section{Framework and Execution Architecture}

The baseline characterization was implemented on the nRF5340DK evaluation kit. The relevant hardware and software elements used for the measurements are summarized below:

\begin{itemize}
    \item \textbf{Application core:} Arm Cortex-M33 configured at 64~MHz, used for the local scalar/ALU and DSP/SIMD convolution measurements.
    \item \textbf{Network core:} Arm Cortex-M33 configured at 64~MHz, used as the second processing element for OpenAMP communication measurements.
    \item \textbf{Asymmetric multiprocessing:} the platform supports AMP execution across the two Cortex-M33 cores, enabling explicit inter-core communication through OpenAMP framework.
    \item \textbf{Zephyr:} provides the embedded RTOS environment, POSIX API support, and OpenAMP support for inter-core communication.
    \item \textbf{Quantized math backend employed by Klepsydra AI:} used because Klepsydra AI supports explicit orchestration across cores, processors, or accelerators, which aligns with the broader scope of offloaded and orchestrated execution.
\end{itemize}

OpenAMP is measured separately from the convolution workload to provide a first reference for communication and data-movement overhead in explicitly orchestrated execution. Two representative inter-core communication patterns are measured:

\begin{itemize}
    \item a one-way \texttt{rpmsg\_send} with a 256-byte payload, which estimates the cost of sending data between cores
    \item a round-trip exchange using a 256-byte payload and a 1-byte acknowledgment, which captures a send-response interaction including synchronization.
\end{itemize}

\section{Results}

This section reports the measured latency of the selected quantized convolution workload and the OpenAMP communication overhead used for orchestration reasoning. The results are organized in three parts. First, the Cortex-M33 measurements are used to compare the constrained scalar baseline against DSP-optimized execution and TensorFlow Lite Micro reference paths. Second, additional measurements are reported on NOEL-V, PolarFire, and ZedBoard to provide reference points for future migration to more capable processing platforms. Third, OpenAMP measurements for data movements between Cortex-M cores.

\subsection{Quantized Convolution Latency}

Table~\ref{tab:cortex_latency} reports the latency of the selected MobileNetV2-derived quantized convolution layer on the nRF5340DK Cortex-M33 application core. The evaluated layer uses an input feature-map size of \(28 \times 28 \times 144\) and a kernel configuration of \(32 \times 1 \times 1 \times 144\). The same workload is executed using the characterized quantized math backend and TensorFlow Lite Micro, each with scalar and optimized execution paths.

\begin{table}[htbp]
\caption{Execution Latency on the Cortex-M33 application core}
\begin{center}
\begin{tabular}{|c|c|c|}
\hline
\textbf{AI Backend} & \textbf{Execution Path} & \textbf{Latency [ms]} \\
\hline
Characterized math backend & ALU & 421.942 \\
\hline
Characterized math backend & DSP & 112.744 \\
\hline
TFLite Micro & ALU & 616.389 \\
\hline
TFLite Micro + CMSIS-NN & DSP & 119.202 \\
\hline
\end{tabular}
\label{tab:cortex_latency}
\end{center}
\end{table}

The DSP speedup is computed as

\begin{equation}
S_{\mathrm{DSP}} =
\frac{T_{\mathrm{ALU}}}{T_{\mathrm{DSP}}}.
\end{equation}

For the characterized math backend, the DSP-optimized path reduces the latency from 421.942~ms to 112.744~ms, corresponding to a speedup of approximately \(3.74\times\) with respect to the scalar/ALU baseline. Under the optimized paths, this math backend achieves approximately \(6\%\) lower latency than the TensorFlow Lite Micro + CMSIS-NN path for the selected workload.

Additional time measurements using the characterized math backend were performed on processing targets beyond the Cortex-M33 baseline. The selected configurations include both space-oriented and commercial processing targets:

\begin{itemize}
    \item \textbf{NOEL-V:} RISC-V architecture, 100~MHz clock frequency, two cores used.
    \item \textbf{PolarFire:} RISC-V architecture, 600~MHz clock frequency, four cores used.
    \item \textbf{ZedBoard:} Arm Cortex-A9 architecture, 667~MHz clock frequency, four cores used.
\end{itemize}

The corresponding measured raw latencies are shown in Table~\ref{tab:platform_latency}.

\begin{table}[htbp]
\caption{Measured Execution Latency Across Evaluated Execution platforms}
\begin{center}
\begin{tabular}{|c|c|c|c|}
\hline
\textbf{Platform} & \textbf{Execution Platform} & \textbf{Path} & \textbf{Latency [ms]} \\
\hline
nRF5340DK & Cortex-M33 app. core & ALU & 421.942 \\
\hline
nRF5340DK & Cortex-M33 app. core & DSP & 112.744 \\
\hline
NOEL-V & RISC-V & ALU & 63.10 \\
\hline
PolarFire & RISC-V & ALU & 6.09 \\
\hline
ZedBoard & Arm Cortex-A9 & SIMD & 1.42 \\
\hline
\end{tabular}
\label{tab:platform_latency}
\end{center}
\end{table}

Because raw latency depends on platform characteristics such as clock frequency and the number of active cores, the measured execution times were normalized to support cross-platform comparison. The normalization applies two adjustments:

\begin{itemize}
    \item \textbf{Clock-frequency adjustment:} execution time is scaled to a reference frequency of 1~GHz to reduce the direct effect of different clock rates.
    \item \textbf{Core-count adjustment:} execution time is normalized to a four-core reference configuration, allowing platforms with different numbers of active cores to be compared using a common metric.
\end{itemize}

\begin{figure}[htbp]
\centerline{\scalebox{0.50}{\includegraphics{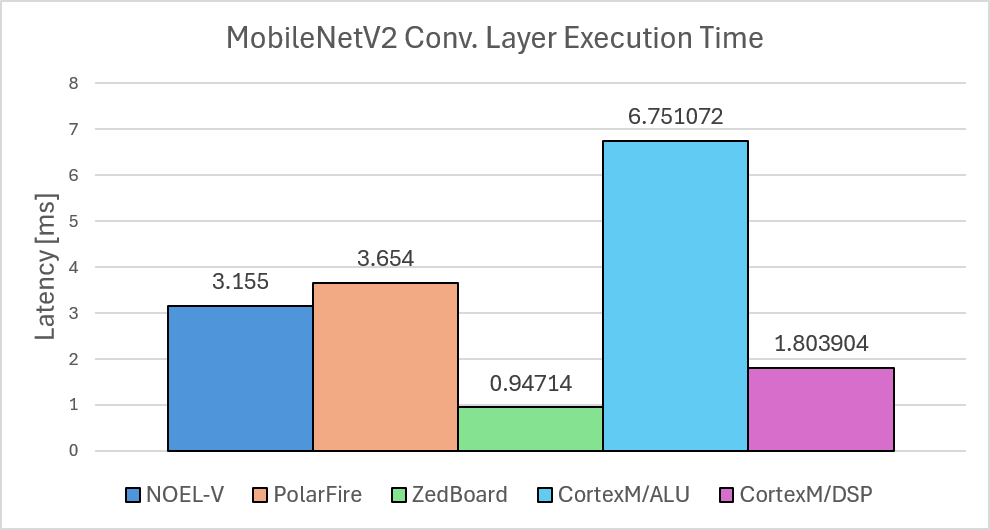}}}
\caption{Normalized Execution Latency Across Evaluated Execution Platforms}
\label{fig:normalized_latency}
\end{figure}

Considering that Cortex-M33 results were normalized considering single core execution and 64 MHz working frequency, the corresponding normalized results are reported in Table~\ref{tab:normalized_latency} and illustrated in Fig.~\ref{fig:normalized_latency}.

\begin{table}[htbp]
\caption{Normalized Execution Latency Across Evaluated Execution Platforms}
\begin{center}
\begin{tabular}{|c|c|c|c|}
\hline
\textbf{Platform} & \textbf{Architecture} & \textbf{Execution Path} & \textbf{Latency [ms]} \\
\hline
NOEL-V & RISC-V & ALU & 3.155 \\
\hline
PolarFire & RISC-V & ALU & 3.654 \\
\hline
ZedBoard & Arm Cortex-A9 & SIMD & 0.947 \\
\hline
nRF5340DK & Cortex-M33 & ALU & 6.751 \\
\hline
nRF5340DK & Cortex-M33 & DSP & 1.804 \\
\hline
\end{tabular}
\label{tab:normalized_latency}
\end{center}
\end{table}

\subsection{OpenAMP Communication Overhead}

Table~\ref{tab:openamp_results} reports the OpenAMP communication measurements between the two Cortex-M33 cores of the nRF5340DK. For these measurements, both the application core and the network core were configured at 64~MHz. The convolution workload is not offloaded in this experiment; instead, these measurements provide a communication reference for explicitly orchestrated execution.

\begin{table}[htbp]
\caption{OpenAMP Communication Measurements Between Cortex-M33 Cores}
\begin{center}
\begin{tabular}{|c|c|c|c|}
\hline
\textbf{Case} & \textbf{Payload} & \textbf{Cycles} & \textbf{Latency [$\mu$s]} \\
\hline
\texttt{rpmsg\_send} & 256 B sent & 2154 & 33.656 \\
\hline
Round-trip & 256 B sent + 1 B ack & 7033 & 109.890 \\
\hline
\end{tabular}
\label{tab:openamp_results}
\end{center}
\end{table}

The difference between the round-trip and one-way measurements is approximately \(76.234~\mu\mathrm{s}\), which represents the additional cost of the response and synchronization path for this interaction pattern.

\section{Conclusions}
This paper presents a baseline characterization of quantized AI execution on a highly resource-constrained embedded platform, together with a methodology for using this baseline to estimate execution latency in more complex inference and orchestration scenarios.

The results show that backend optimization has a significant impact on quantized inference latency. For the selected workload, the DSP/SIMD-optimized path of the characterized math backend provided a speedup of approximately \(3.74\times\) over the scalar/ALU baseline. This result is consistent with the implemented DSP operations, which exploits 32-bit registers to operate on packed 8-bit quantized data. Since four 8-bit values can be represented within a 32-bit word, the optimized path can approach a four-way improvement over scalar execution for the arithmetic-dominated part of the convolution. The measured speedup remains below the ideal \(4\times\) value because the total latency also includes memory access, loop control, accumulation, requantization, saturation, and backend invocation overhead. In addition, the execution time results obtained from the scalar/ALU and DSP/SIMD paths confirm that backend implementation affects latency even when the quantized operation and platform are fixed.

The measured convolution latency also provides the basis for deriving a backend-specific time-per-MAC indicator when combined with the MAC count of the selected layer. This indicator is intended as a planning-level KPI for estimating the latency of other convolution layers executed under the same framework-backend environment. It reflects the combined effect of the platform, backend implementation, memory-access behavior, data layout, and invocation overhead.

The additional raw and normalized comparisons provide reference points for positioning the Cortex-M33 baseline relative to both space-oriented and commercial processing targets. The raw results reflect overall platform-scale differences, such as clock frequency and core count, while the normalized results highlight the differences related to architecture, backend implementation, and data-movement efficiency.

Finally, the OpenAMP measurements provide the communication term required for future orchestration analysis. The measured one-way and round-trip communication times are small compared with the local convolution latency, but they remain relevant when offloading is applied repeatedly or at fine granularity. For this reason, the proposed baseline supports both computation-time estimation and early evaluation of whether multicore, multi-processor, or accelerator-based execution is likely to provide a net timing benefit.

\section{Future Work}

Future work will apply the proposed baseline methodology to more complex explicitly orchestrated AI deployments on Cortex-M-class constrained platforms. In particular, the MAC-based timing indicator derived from the measured convolution baseline will be used to estimate the execution time of complete inference pipelines on the same class of devices. These estimates will then be combined with measured communication and synchronization overheads to evaluate which offloading or parallel-execution configurations across Cortex-M processing elements are likely to provide a net timing benefit.

In addition, these estimates will be used as a plausibility check after implementation. Large deviations between estimated and measured execution time can reveal additional overheads such as data-transfer bottlenecks, synchronization delay, load imbalance, or framework-level integration costs.

After validation on the constrained Cortex-M platform, the same methodology can be extended toward more space-oriented processing targets, including LEON/NOEL-V processors and FPGA-based acceleration, where communication mechanisms and architectural constraints differ from those evaluated in this work.

\section*{Acknowledgment}

We thank Innosuisse for the financial support of this research under the umbrella of the ARITHMETIC project, Innosuisse project No.~121.834 IP-ENG.

\end{document}